\documentclass[aps,prb,10pt,twocolumn,floatfix,footinbib,superscriptaddress,longbibliography,nobibnotes]{revtex4-2}

\usepackage{amsmath}
\usepackage{amssymb}

\usepackage{mathtools}
\usepackage{wasysym}
\usepackage[T1]{fontenc}
\usepackage{amsfonts}
\usepackage{newtxtext}
\usepackage[varvw]{newtxmath}
\usepackage{dsfont}                 
\usepackage{bbold}                  
\usepackage[normalem]{ulem}         

\usepackage{enumerate}
\usepackage[shortlabels]{enumitem}
\usepackage{footmisc}

\usepackage[dvipsnames,table]{xcolor}
\usepackage{graphicx}
\graphicspath{{figures/}} 

\usepackage{physics}   
\usepackage{csquotes}  
\usepackage{comment}   
\usepackage{relsize}   

\usepackage[caption=false]{subfig}
\usepackage[colorlinks=true,citecolor=cyan,linkcolor=magenta,filecolor=magenta,bookmarksnumbered]{hyperref}
\usepackage[capitalize]{cleveref}

\usepackage{xcolor}

\usepackage{orcidlink}

\newcommand{\RNum}[1]{\uppercase\expandafter{\romannumeral #1\relax}}

\makeatletter
\newif\ifnatc@aftersection
\natc@aftersectionfalse

\newcommand{\sectitle}[1]{%
  \par\addvspace{\medskipamount}%
  \noindent\phantomsection
  {\large\bfseries #1\par}%
  \@afterindentfalse\@afterheading
  \global\natc@aftersectiontrue
  \nopagebreak
}

\newcommand{\subsectitle}[1]{%
  \par\addvspace{\ifnatc@aftersection 0pt\else \medskipamount\fi}%
  \noindent\phantomsection
  {\bfseries #1\par}%
  \@afterindentfalse\@afterheading
  \global\natc@aftersectionfalse
  \nopagebreak
}

\makeatother

\begin{document}

\title{Kitaev spin liquid in superconducting networks}

\author{Guilherme Delfino}

\author{Mehmet Dede}
\affiliation{Department of Physics and Astronomy, Purdue University, West Lafayette, IN 47907}

\author{Dmitry Green}
\affiliation{Department of Physics, Boston University, Boston, MA 02215}
\affiliation
{AppliedTQC, New York, NY 10065}

\author{\\Michael J. Manfra}
\affiliation{Department of Physics and Astronomy, Purdue University, West Lafayette, IN 47907}
\affiliation{Elmore Family School of Electrical and Computer Engineering, Purdue University, West Lafayette, IN 47907}
\affiliation{School of Materials Engineering, Purdue University, West Lafayette, IN 47907}
\affiliation{Purdue Quantum Science and Engineering Institute, Purdue University, West Lafayette, IN 47907}
\affiliation{Microsoft Quantum, West Lafayette, IN 47907}

\author{Charles M. Marcus}
\affiliation{Episteme, San Francisco, CA, 94103}
\affiliation{Center for Quantum Devices, Niels Bohr Institute, University of Copenhagen, Copenhagen 2100,
Denmark}
\affiliation{Materials Science and Engineering, and Department of Physics, University of Washington,
Seattle, WA 98195}

\author{Claudio Chamon}
\affiliation{Department of Physics and Astronomy, Purdue University, West Lafayette, IN 47907}
\affiliation{Purdue Quantum Science and Engineering Institute, Purdue University, West Lafayette, IN 47907}


\begin{abstract}

 We propose a realization of the Kitaev honeycomb Hamiltonian -- an archetypal spin-liquid model -- in a superconducting metamaterial. The architecture consists of Cooper-pair boxes coupled through depleted semiconductor--superconductor heterostructures that do not require spin--orbit coupling. The Cooper-pair boxes encode effective spin degrees of freedom, while normal and anomalous virtual propagation through the heterostructures mediate bond-directional interactions. Two key control parameters are an out-of-plane magnetic flux and the semiconductor Fermi energy. The former controls interference and distinguishes the bond directions, while tuning the latter close to the bottom of the band gives rise to an emergent Nambu-exchange symmetry that enforces the required bond directionality.  Through numerical calculations, we identify an operating regime with controlled corrections, with associated energy and length scales within experimental reach. These results establish a route toward equilibrium quantum spin liquids in engineered superconducting networks.
\end{abstract}

\maketitle


\section{Introduction}
Realizing quantum spin liquids in controlled experimental settings remains a central challenge in condensed matter physics. Efforts toward this goal have coalesced into three distinct experimental paradigms.
The first approach involves searches for quantum spin liquid behavior in naturally occurring strongly correlated materials~\cite{Jackeli2009kitaev,Trebst2022kitaev,Matsuda2025kitaev,Kim2022alpha, cai2025quantum, Nordlander2025signatures}. Despite progress, evidence for equilibrium spin-liquid phases remains inconclusive, owing to limited microscopic control and competing ordered states.
A second route leverages programmable quantum processors to digitally or analogically simulate topologically ordered states~\cite{Lo2026universal,Xu2024nonabelian,Iqbal2025qutrit,Minev2025realizing,Google2023non,Xu2023digital, Semeghini2021probing,Bornet2026diracspinliquidcandidate, Gammon2026simulating}. These platforms offer remarkable tunability and controlled access to topological states, but they typically rely on dynamical state preparation rather than realizing a static Hamiltonian's ground state.
Here, we pursue a complementary third route: engineering synthetic quantum metamaterials designed to possess intrinsic topological order at the hardware level. This approach preserves the notion of equilibrium phases of matter while retaining microscopic control over interactions and energy scales, overcoming key limitations of naturally occurring materials.

Theoretical~\cite{Rosdahl2018rectifier, Sau2012realizing, Leijnse2012poorman, Fulga2013adaptive}  and experimental~\cite{Lutchyn2018majorana, Cai2019magnon, Aghaee2025distinct, Microsoft2025interferometric, Microsoft2023InAs, Bordin2025enhanced, Bordin2023tunable, Bordin2024crossed} efforts have explored superconducting platforms to realize one-dimensional chains hosting Majorana zero modes.
 Superconducting architectures have also been proposed for realizing two-dimensional fermionic topological phases \cite{Li2016two, Zang2021competing, Maiani2021topological, Liu2022optimizing, Delfino2025splitters}, as well as bosonic topological order based purely on Cooper-pair degrees of freedom~\cite{Ioffe2002Possible,Ioffe2003topological, Vidal2004nonabelian, Gladchenko2009superconducting, Sameti2017superconducting, Chamon2020qsl_combinatorial, Green2023abelian_qsl, Yang2021quantum_double_z3}. These developments showcase the versatility of superconducting platforms and motivate the search for architectures in which a controlled low-energy Hamiltonian supports equilibrium topological phases.

Here, using only bosonic (Cooper-pair) degrees of freedom, we construct a network whose leading low-energy Hamiltonian realizes the Kitaev model on the honeycomb lattice. Spin-$1/2$ degrees of freedom, represented by Pauli operators $\hat\sigma_i^{\mathrm a}$ at each vertex $i$, are coupled through bond-directional interactions along $\mathrm{x}$, $\mathrm{y}$, and $\mathrm{z}$ links~\cite{Kitaev2006exactly},
\begin{eqnarray}
\widehat H_{\mathrm{Kitaev}}
=
-\sum_{\mathrm a=\mathrm{x,y,z}}
J_{\mathrm a}
\sum_{\langle i,j\rangle\in\mathrm{a-links}}
\hat\sigma_i^{\mathrm a}\hat\sigma_j^{\mathrm a}.
\label{kitaev}
\end{eqnarray}
The model supports a gapless spin-liquid phase near the isotropic point, \(J_{\mathrm{x}}\simeq J_{\mathrm{y}}\simeq J_{\mathrm{z}}\), as well as gapped Abelian phases in the strongly anisotropic regime. Our platform primarily targets the approximately isotropic limit and the associated gapless Kitaev spin liquid. When supplemented by suitable time-reversal-breaking single-spin fields, the system can also enter a gapped non-Abelian phase characterized by chiral Majorana edge modes and Ising anyons~\cite{Kitaev2006exactly}.

In our construction, the effective two-level systems are encoded in the even-electron-number states of small superconducting islands, protected from the odd-parity sector by a parity gap. Bond-directional interactions between these effective spins are mediated by depleted semiconductor--superconductor heterostructures through Cooper-pair cotunneling ($\mathrm{ECT}^2$) and double crossed Andreev reflection ($\mathrm{CAR}^2$). These processes can be viewed as Cooper-pair counterparts of single-electron elastic cotunneling and crossed Andreev reflection~\cite{Falci2001correlated,Melin2002transport,Feinberg2003andreev,Morten2006circuit, Leijnse2013coupling, Reeg2017destructive}. 
Their relative strength can be controlled by depleting the semiconductor close to the bottom of its band. At the band edge, the leading low-energy $\mathrm{ECT}^2$ and $\mathrm{CAR}^2$ contributions become equal, yielding the desired bond-directional interaction without relying on spin--orbit coupling. We trace this matching to an emergent Nambu-exchange symmetry of the lowest energy states manifold, which remains approximately valid over a finite range of carrier densities.

\begin{figure*}[!t]
    \centering
\includegraphics[width=0.8\linewidth]{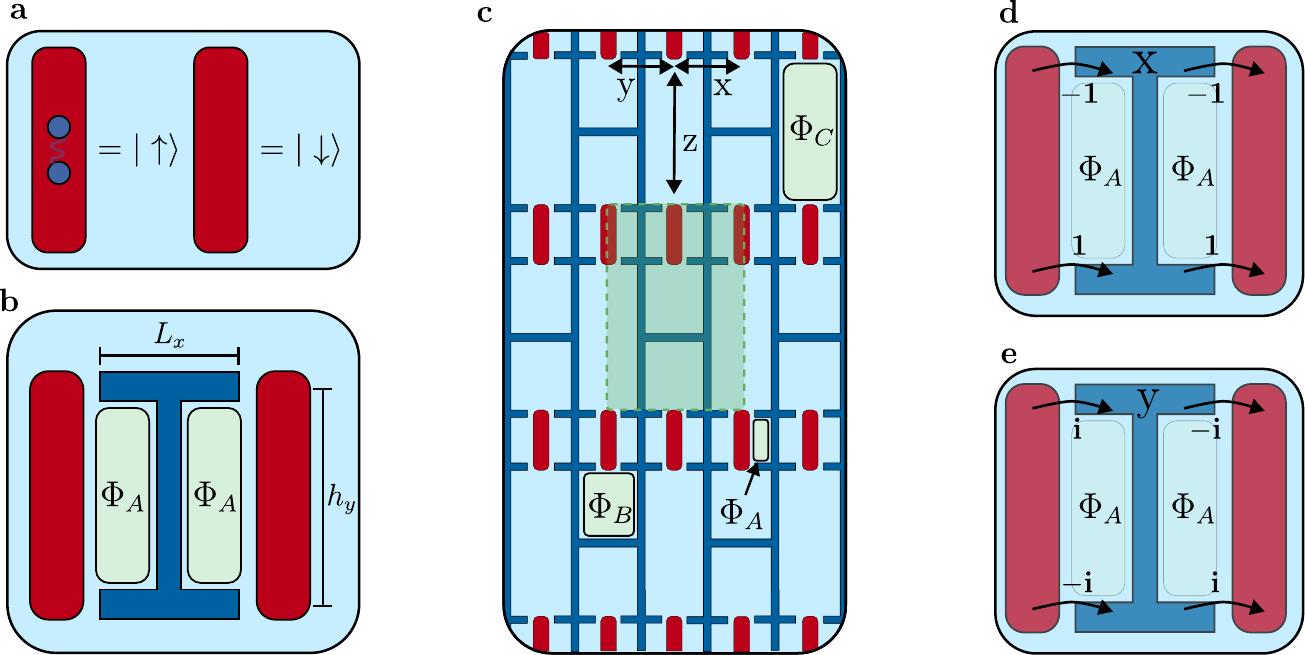}
\caption{\textbf{a} Cooper-pair boxes (red) realize effective spin-$1/2$ degrees of freedom, with spin up (down) corresponding to the doubly occupied (empty) state. \textbf{b} I-gadget (blue) formed by a proximitized semiconductor element  with horizontal length $L_x$ and vertical leg separation $h_y$, connecting two Cooper-pair boxes. Each side of the I-gadget encloses a magnetic flux $\Phi_A$. \textbf{c} Brick-wall-like network of Cooper-pair boxes coupled through interconnected I-gadgets, with bonds labeled $\mathrm{x}$, $\mathrm{y}$, and $\mathrm{z}$. The green rectangle marks a unit cell consisting of one half-brick and two quarter-bricks. Each quarter-brick encloses a flux $\Phi_B$, while the half-brick encloses $\Phi_C=2\Phi_B$. \textbf{d} Real and \textbf{e} imaginary Peierls phases of the pair-hopping amplitudes along the $\mathrm{x}$ and $\mathrm{y}$ bonds, respectively. These phases implement the required plaquette fluxes and the relative $\pi$ phase between the two ends of each I-gadget.}
    \label{array}
\end{figure*}


\section{Results}
\subsection{Elementary building blocks}

The proposed network consists of two types of elements: small superconducting vertex elements (red in Fig.~\ref{array}a), and superconducting segments shaped as ``I''s, which we refer to as I-gadgets (blue in Fig.~\ref{array}b). While the vertex elements provide two-state degrees of freedom, the edge I-gadgets mediate bond-directional interactions.

The vertex elements operate in a regime where the pairing energy exceeds the Coulomb energy~\cite{Tuominen1997parity,Lafarge1993parity}, such that the empty and doubly occupied configurations are the only accessible states below the parity gap. They can therefore be regarded as Cooper-pair boxes (CPBs)~\cite{Nakamura1999box,Bouchiat1998quantum}, realizing effective two-level systems at the vertices (Fig.~\ref{array}a). We initially assume these two states to be degenerate, postponing the effects of charge bias to the experimental considerations (Sec.~\ref{sec: exp considerations}). Let $\hat\sigma_i^+$ and $\hat\sigma_i^-$ denote the creation and annihilation operators of a Cooper pair on CPB $i$, and let $\hat\sigma_i^z$ distinguish between the empty and doubly occupied states. Restricted to the even-parity subspace, these operators obey the $\mathrm{SU}(2)$ spin algebra and provide the mapping between the electronic and effective spin-$1/2$ degrees of freedom used throughout. An explicit representation in terms of electron operators is given in Sec.~\ref{sec: 
further 
details}.

The link elements, the I-gadgets (Fig.~\ref{array}b), mediate interactions between the CPBs' effective spins through the propagation of virtual quasiparticles, setting a natural horizontal length $L_x$. Their top and bottom legs are separated by $h_y$. In a perpendicular magnetic field, each side of an I-gadget encloses a flux $\Phi_A=\Phi_0/2\ \mathrm{mod}\ \Phi_0$, where $\Phi_0=hc/2e$ is the superconducting flux quantum. At this flux, pair hopping amplitudes associated with the top and bottom legs interfere destructively, canceling unwanted local contributions.

\subsection{Connecting elements}

The system consists of Cooper-pair boxes and I-gadgets arranged in the brick-wall-like geometry of Fig.~\ref{array}c, which is lithographically equivalent to the honeycomb lattice with $\mathrm{x}$, $\mathrm{y}$, and $\mathrm{z}$ bonds. The I-gadgets are interconnected by superconducting segments and can therefore be treated as a single extended condensate with negligible charging energy. The $\mathrm{x}$ and $\mathrm{y}$ bonds are mediated directly by the superconducting network, whereas the $\mathrm{z}$ bonds connect neighboring boxes without an intermediate superconducting segment.

The device geometry is designed such that its elementary loops enclose prescribed magnetic fluxes. In addition to $\Phi_A$ introduced above, the construction requires $\Phi_B=\Phi_0/4\ \mathrm{mod}\ \Phi_0$ through each quarter-brick loop. It is responsible for producing the alternating interactions along the $\mathrm{x}$ and $\mathrm{y}$ bonds. The half-brick loop shown in Fig.~\ref{array}c, on the other hand, encloses $\Phi_C=2\Phi_B$ and imposes no additional constraint. All required fluxes can, in principle, be generated by a single uniform perpendicular magnetic field, with their commensurability encoded in the device geometry. The resulting dynamics is governed by local link Hamiltonians whose form depends on whether neighboring CPB vertices $i$ and $j$ are connected by an $\mathrm{x}$, $\mathrm{y}$, or $\mathrm{z}$ bond.

\subsection{Bond-directional interactions}

A key distinction of the $\mathrm{z}$ edges is the absence of an intermediate superconducting segment (Fig.~\ref{array}c) and, consequently, of the associated electrostatic screening. Coulomb interactions between the neighboring CPBs therefore remain appreciable across these bonds. The corresponding vertices interact capacitively through a density--density coupling,
\begin{eqnarray}
    \widehat H_{ij}^{\mathrm{z}}
    &=& U_{ij}\,\hat\sigma_i^{\mathrm z}\hat\sigma_j^{\mathrm z}+g_{i}^z\hat \sigma_i^z+g_{j}^z\hat \sigma_j^z.
    \label{eq: insulating strip}
\end{eqnarray}
The linear terms $g_k^z$ correspond to charge biases. Provided they remain sufficiently small, individual tuning of each CPB is not required and residual fields may even be exploited to access the gapped non-Abelian phase, as discussed in the experimental considerations (Sec.~\ref{sec: exp considerations}).
The resulting two-body $\hat\sigma_i^{\mathrm z}\hat\sigma_j^{\mathrm z}$ interaction thus represents the effective capacitive coupling between vertices $i$ and $j$~\footnote{The sign of the $\hat\sigma_i^{\mathrm z}\hat\sigma_j^{\mathrm z}$ interaction can be reversed by the finite-depth local unitary $\prod_{i\in\mathrm{even\,rows}}\hat\sigma_i^{\mathrm x}$, which acts on every other row of the brick-wall-like lattice without affecting the topological physics.}.

Next, we consider neighboring spins connected by the horizontal I-gadgets, which define the $\mathrm{x}$ and $\mathrm{y}$ edges. As shown below, these links mediate the bond-directional interactions $\hat\sigma_i^{\mathrm{x}}\hat\sigma_j^{\mathrm{x}}$ and $\hat\sigma_i^{\mathrm{y}}\hat\sigma_j^{\mathrm{y}}$ through exchange and anomalous pair-transfer processes. To derive these interactions, let $\hat b_{i,p}^{\dagger}$ denote a Cooper-pair creation operator localized near the $i$-th end of leg $p$ of the I-gadget, with $p=t,b$ labeling the top and bottom legs respectively,
\begin{eqnarray}
    \hat b_{i,p}^{\dagger}
    &=&
    \int \mathrm d^2\mathbf r\,
    f_{i,p}(\mathbf r)\,\hat b^\dagger(\mathbf r).
\end{eqnarray}
Here, $\hat b^\dagger(\mathbf r)$ creates a Cooper pair at position $\mathbf r$ in the proximitized semiconductor, while $f_{i,p}(\mathbf r)$ is an envelope function exponentially localized near $(i,p)$ normalized by $  \int \mathrm d^2 \mathbf r \, \left|f_{i,p}(\mathbf r)\right|^2=1.$

\begin{figure*}[t]
    \centering
    \includegraphics[width=0.7\linewidth]{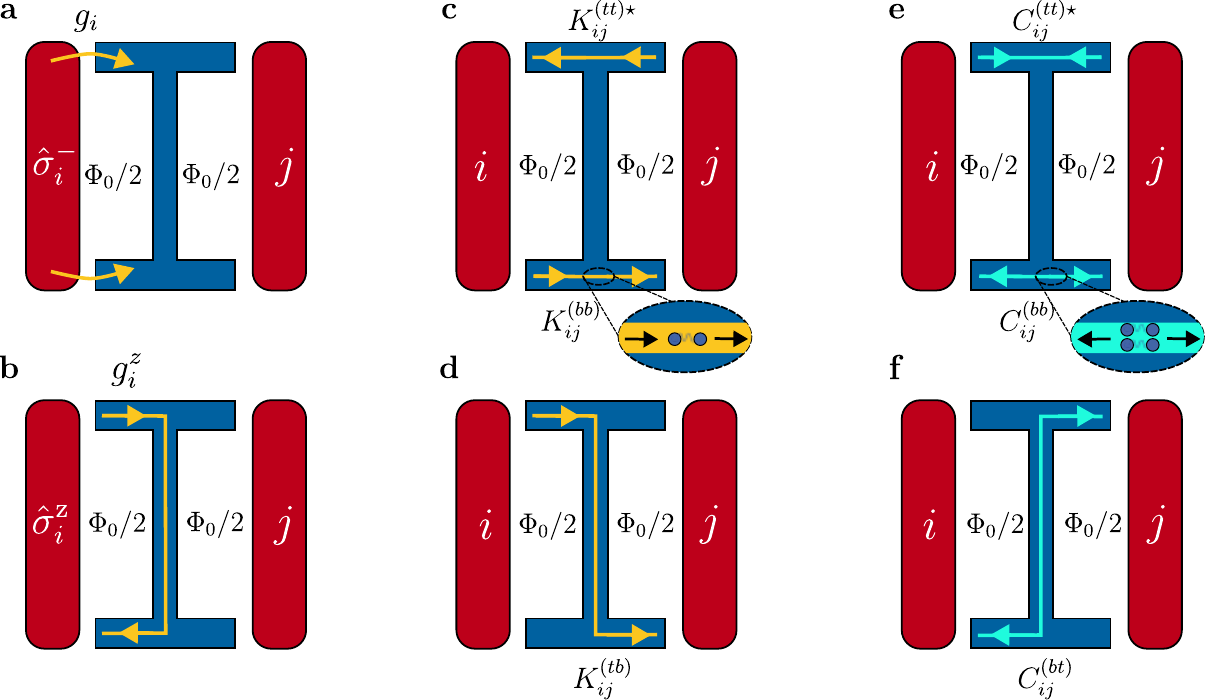}
    \caption{ Virtual processes in a superconducting I-gadget connecting Cooper-pair boxes $i$ and $j$ in the presence of flux $\Phi_A=\Phi_0/2\ \mathrm{mod}\ \Phi_0$. \textbf{a} Pair hopping amplitudes associated with the top and bottom legs interfere destructively, yielding $g_i=0$. \textbf{b} Closed virtual process in which a Cooper pair leaves and returns to the same box, generating a local longitudinal field $g_i^z\hat\sigma_i^{\mathrm z}$. \textbf{c} Same-leg $\mathrm{ECT}^2$ processes (yellow): Cooper-pair transfer from $i$ to $j$ along the bottom leg, $K_{ij}^{(bb)}$, and the reverse process along the top leg, $K_{ij}^{(tt)\star}$. \textbf{d} Cross-leg $\mathrm{ECT}^2$ process $K_{ij}^{(tb)}$. \textbf{e} Same-leg $\mathrm{CAR}^2$ processes (green):  anomalous pair transfer along the bottom leg, $C_{ij}^{(bb)}$, and the reverse process along the top leg, $C_{ij}^{(tt)\star}$. \textbf{f} Cross-leg $\mathrm{CAR}^2$ process $C_{ij}^{(bt)}$. Arrows indicate the direction of Cooper-pair transfer and reversing them corresponds to the complex-conjugate process. Cross-leg contributions sample the vertical bridge and are generally smaller than same-leg contributions in the elongated geometry considered here.}
    \label{fig: transport processes}
\end{figure*}

Each leg couples the I-gadget to its neighboring CPBs $i$ and $j$ through
\begin{eqnarray}
    \widehat V_{ij}^{(p)}
    &=&
    \gamma_{i,p}\,\hat b_{i,p}^{\dagger}\hat\sigma_i^-
    +\gamma_{j,p}\,\hat b_{j,p}^{\dagger}\hat\sigma_j^-
    +\mathrm{h.c.},
    \label{eq: coupling}
\end{eqnarray}
which describes Cooper-pair hopping between the CPBs and the I-gadget. The $\gamma_{i,p}$ are microscopic pair-hopping amplitudes whose magnitudes are controlled primarily by the contact-barrier thickness, while their Peierls phases are set by the magnetic fluxes $\Phi_A$ and $\Phi_B$, as shown explicitly in Fig.~\ref{array}d,e.

The effective theory for the CPBs obtained by integrating out the superconducting elements contains four classes of processes, illustrated in Fig.~\ref{fig: transport processes}. Local hopping  transfers Cooper pairs between a CPB and the adjacent I-gadget, generating $g_i^\star\hat\sigma_i^+ + g_i\hat\sigma_i^-$ (Fig.~\ref{fig: transport processes}a). Virtual processes in which a Cooper pair leaves and returns to the same CPB instead generate local fields $g_i^z\hat\sigma_i^{\mathrm z}$ (Fig.~\ref{fig: transport processes}b). These fields decay exponentially with the height of the I-gadget and are negligible for the parameters considered here, and we therefore omit them below. We defer the explicit expression for the corresponding decay length to Eq.~\eqref{eq: edge coherence length}.

Processes connecting neighboring CPBs give rise to the two interactions central to our construction. Quasiparticle-mediated exchange (ECT$^2$) transfers a Cooper pair from one CPB to the other, generating $K_{ij}\hat\sigma_i^-\hat\sigma_j^+
+K_{ij}^\star\hat\sigma_i^+\hat\sigma_j^-,$
whereas anomalous pair transfer (CAR$^2$), mediated by the superconducting condensate, creates or removes Cooper pairs on both CPBs $C_{ij}\hat\sigma_i^+\hat\sigma_j^+
+C_{ij}^\star\hat\sigma_i^-\hat\sigma_j^-$.
Both amplitudes receive contributions from the top (t) and bottom (b) legs of the I-gadget,
\begin{subequations}\label{eq: physical H totals}
\begin{eqnarray}
K_{ij} &=& K_{ij}^{(tt)}+K_{ij}^{(bb)}-K_{ij}^{(tb)}-K_{ij}^{(bt)},\\
C_{ij} &=& C_{ij}^{(tt)}+C_{ij}^{(bb)}-C_{ij}^{(tb)}-C_{ij}^{(bt)},
\end{eqnarray}
\end{subequations}
where $K_{ij}^{(pq)}$ and $C_{ij}^{(pq)}$ denotes processes between the $i$-th end of leg $p$ and the $j$-th end of leg $q$, with $p,q=t,b$ [Fig.~\ref{fig: transport processes}c--f]. The relative signs of these contributions arise from the Peierls phases associated with the flux $\Phi_A$ (Fig.~\ref{array}d,e).

As illustrated in Fig.~\ref{array}d,e, the $\mathrm{x}$ and $\mathrm{y}$ links carry purely real and purely imaginary Cooper-pair hopping phases, respectively~\footnote{Although this representation is gauge dependent, a different gauge choice would lead to an equivalent Kitaev-like interaction expressed in a different spin basis}. The distinction between these link types follows from the presence of magnetic fluxes $\Phi_B$ and $\Phi_C$. Extracting the superconducting and contact phases so that $g_i$, $K_{ij}$, and $C_{ij}$ are real, the resulting interactions are
\begin{subequations}
\begin{widetext}
\begin{align}
\widehat H_{ij}^{\mathrm{x}}
&=-\sum_{k={i,j}}g_k\left( \hat\sigma_k^-+ \hat\sigma_k^+\right)
-\left[
K_{ij}
\left(
\hat{\sigma}_i^{+}\hat{\sigma}_j^{-}
+
\hat{\sigma}_i^{-}\hat{\sigma}_j^{+}
\right)
+
C_{ij}
\left(
\hat{\sigma}_i^{+}\hat{\sigma}_j^{+}
+
\hat{\sigma}_i^{-}\hat{\sigma}_j^{-}
\right)
\right] ,
\label{eq: processes x}
\\
\widehat H_{ij}^{\mathrm{y}}
&=\sum_{k={i,j}}g_k\left(\mathrm{i}\, \hat\sigma_k^--\mathrm{i}\, \hat\sigma_k^+\right)
-\left[
K_{ij}
\left(
\hat{\sigma}_i^{+}\hat{\sigma}_j^{-}
+
\hat{\sigma}_i^{-}\hat{\sigma}_j^{+}
\right)
-
{C_{ij}}
\left(
\hat{\sigma}_i^{+}\hat{\sigma}_j^{+}
+
\hat{\sigma}_i^{-}\hat{\sigma}_j^{-}
\right)
\right] .
\label{eq: processes y}
\end{align}
\end{widetext}
\end{subequations}
The relative minus sign between ECT$^2$ and CAR$^2$ on the $\mathrm{y}$ links follows directly from the imaginary hopping phases. In a CAR$^2$ process the two Cooper pairs propagate toward opposite CPBs, reversing one of the directed phases in Fig.~\ref{array}e and hence its sign. No such sign change occurs for the real hopping phases on the $\mathrm{x}$ links [Fig.~\ref{array}d].

 It is instructive to rewrite these link Hamiltonians in the usual Pauli basis. Define $J_{ij}\equiv{(K_{ij}+C_{ij})}/{2}$
and
$\alpha_{ij}\equiv{(K_{ij}-C_{ij}})/2$, then
\begin{subequations}
\begin{align}
\widehat H_{ij}^{\mathrm{x}}
&=
\sum_{k=i,j}g_k\,\hat\sigma_k^{\mathrm{x}}
-
J_{ij}\hat\sigma_i^{\mathrm{x}}\hat\sigma_j^{\mathrm{x}}
-
\alpha_{ij}\hat\sigma_i^{\mathrm{y}}\hat\sigma_j^{\mathrm{y}},
\label{eq: effective_SC x}
\\
\widehat H_{ij}^{\mathrm{y}}
&=
\sum_{k=i,j}g_k\,\hat\sigma_k^{\mathrm{y}}
-
J_{ij}\hat\sigma_i^{\mathrm{y}}\hat\sigma_j^{\mathrm{y}}-
\alpha_{ij}\hat\sigma_i^{\mathrm{x}}\hat\sigma_j^{\mathrm{x}}.
\label{eq: effective_SC y}
\end{align}
\end{subequations}
Purely bond-directional interactions emerge when both $g_k = 0$ and $K_{ij}=C_{ij}$. In this limit, the undesired terms proportional to both $g_k$ and $\alpha_{ij}$ vanish, leaving $-J_{\mathrm{x}}\hat\sigma_i^{\mathrm{x}}\hat\sigma_j^{\mathrm{x}}$ on $\mathrm{x}$ bonds and $-J_{\mathrm{y}}\hat\sigma_i^{\mathrm{y}}\hat\sigma_j^{\mathrm{y}}$ on $\mathrm{y}$ bonds, with $J_{\mathrm{x},\mathrm{y}}=J_{ij}$.

In the following, we derive the coefficients $g_i$, $K_{ij}$, and $C_{ij}$ using a Schrieffer--Wolff transformation and establish the conditions under which the effective model reduces to the Kitaev Hamiltonian. Before proceeding with the derivation, we state the central result: in the low electron-density limit of small I-gadgets $K_{ij}=C_{ij}$. This equality follows from a unitary Nambu-exchange symmetry that is exact within the lowest-mode sector and emerges approximately in the full I-gadget.

\subsection{Local pair tunneling, ECT$^2$ and CAR$^2$}

Let $|\Omega\rangle$ denote the ground state of the I-gadget Hamiltonian $\widehat H_{ij}^{\mathrm{SC}}$ between CPBs $i$ and $j$. The effective Hamiltonians in Eqs.~\eqref{eq: effective_SC x} and~\eqref{eq: effective_SC y} are obtained perturbatively through a Schrieffer--Wolff transformation. At first order, local pair hopping processes generate
\begin{eqnarray}
g_i
\equiv
|\gamma_{i,t}|
\left\langle\Omega\left|\hat b_{i,t}^\dagger\right|\Omega\right\rangle
-
|\gamma_{i,b}|
\left\langle\Omega\left|\hat b_{i,b}^\dagger\right|\Omega\right\rangle .
\end{eqnarray}
For an I-gadget symmetric under exchange of its top and bottom legs, $|\gamma_{i,t}|=|\gamma_{i,b}|$ and $\langle\Omega|\hat b_{i,t}^\dagger|\Omega\rangle=\langle\Omega|\hat b_{i,b}^\dagger|\Omega\rangle$, so that $g_i=0$. Fabrication imperfections generally break this symmetry and generate residual local fields, which remain perturbative provided the leg asymmetry is small, $(|\gamma_{i,t}|-|\gamma_{i,b}|)/|\gamma_{i,t}|\ll1$. Together with the longitudinal fields $g_i^z$, these residual transverse fields can break time-reversal symmetry and open the Majorana gap required to access the non-Abelian phase. These processes set the local Josephson coupling and allow us to define a measurable Josephson energy characterizing Cooper-pair hopping at each contact,
\begin{eqnarray}
E_{J,i,p}
\equiv
|\gamma_{i,p}|
\left|
\langle\Omega|\hat b_{i,p}^{\dagger}|\Omega\rangle
\right|,
\label{eq: local Josephson energy}
\end{eqnarray}
which we use below to eliminate the bare coupling amplitudes $|\gamma_{i,p}|$.

The remaining interactions arise at second order. We define the superconducting propagator 
\begin{equation*}
\widehat G_{ij}^{\mathrm{SC}}
\equiv
\widehat Q^{\mathrm{SC}}
\left(\widehat H_{ij}^{\mathrm{SC}}-E_0^{\mathrm{SC}}\right)^{-1}
\widehat Q^{\mathrm{SC}},
\end{equation*}
where $E_0^{\mathrm{SC}}$ is the ground-state energy of $\widehat H_{ij}^{\mathrm{SC}}$ and $\widehat Q^{\mathrm{SC}}=1-|\Omega\rangle\langle\Omega|$ projects onto the subspace orthogonal to $|\Omega\rangle$. The corresponding $\mathrm{ECT}^2$ and $\mathrm{CAR}^2$ coefficients are
\begin{subequations}
\begin{align}
K_{ij}^{(pq)}
=
|\gamma_{i,p}\gamma_{j,q}|
\left\langle\Omega\left|
\hat b_{i,p}^{\dagger}\,
\widehat G_{ij}^{\mathrm{SC}}\,
\hat b_{j,q}+\hat b_{j,q}\,
\widehat G_{ij}^{\mathrm{SC}}\,\hat b_{i,p}^{\dagger}
\right|\Omega\right\rangle,\label{correlations K}\\
C_{ij}^{(pq)}
=
|\gamma_{i,p}\gamma_{j,q}|
\left\langle\Omega\left|
\hat b_{i,p}\,
\widehat G_{ij}^{\mathrm{SC}}\,
\hat b_{j,q} + \hat b_{j,q}\,
\widehat G_{ij}^{\mathrm{SC}}\,\hat b_{i,p}
\right|\Omega\right\rangle
.
\label{correlations C}
\end{align}
\end{subequations}
In the above, the two terms in $K_{ij}^{(pq)}$ correspond to pair removal and pair addition in the intermediate superconducting state. The two terms in $C_{ij}^{(pq)}$ are the two possible time orderings of the anomalous process.

\begin{figure}[t]
    \centering
    \includegraphics[width=1.0\linewidth]{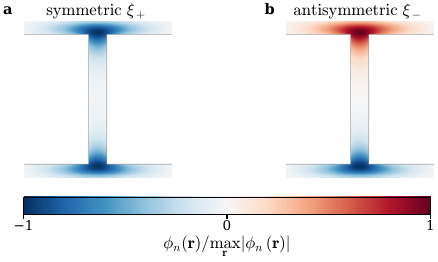}
    \caption{Signed amplitudes of the lowest normal-state mirror-symmetry doublet $\xi_\pm\simeq 0$ of the I-gadget. \textbf{a} Symmetric mode with normal-state energy $\xi_+$ and \textbf{b} Antisymmetric mode with normal-state energy $\xi_-$. The relative sign between the top and bottom legs distinguishes the two sectors. The energy splitting between these two states is exponentially small in the I-gadget height.}
    \label{fig: H lowest parity modes}
\end{figure}

 Cross-leg processes ($p\neq q$) are sensitive to symmetric and antisymmetric combinations of the wavefunctions across the vertical bridge (Fig.~\ref{fig: H lowest parity modes}) and are exponentially suppressed with the I-gadget height relative to processes confined to a horizontal leg ($p=q$). Again, we defer the explicit expression for the corresponding decay length to Eq.~\eqref{eq: edge coherence length}. A similar mechanism suppresses additional processes enabled by the interconnected superconducting network, since couplings between CPBs beyond nearest neighbors require propagation over longer distances and are therefore exponentially suppressed. We neglect these longer-range contributions in the effective model.

\subsection{ECT$^2$ equals CAR$^2$ at the band edge}

While the $\hat\sigma_i^{\mathrm z}\hat\sigma_j^{\mathrm z}$ interactions along the $\mathrm z$ edges already have the desired form, realizing purely bond-directional interactions along the $\mathrm x$ and $\mathrm y$ edges requires balancing the $\mathrm{ECT}^2$ and $\mathrm{CAR}^2$ amplitudes. This balance can be controlled electrostatically through the semiconductor gate voltage~\cite{Recher2001andreev,Brauer2010nonlocal,Rosdahl2018rectifier,Bordin2023tunable,Feng2025long,Bordin2025enhanced,Bordin2024crossed,Fulop2015splitter,Schindele2012nearunit,Wang2022singlet}, which tunes the Fermi energy $E_{\mathrm F}$ measured from the bottom of the normal-state band. We focus on the low-density regime, where the semiconductor is strongly depleted while the parent superconductor remains a large condensate with negligible charging energy.

Consider the real, normalized eigenfunctions $\phi_n(\mathbf r)$  of the normal-state Hamiltonian $h_0\phi_n(\mathbf r)
=
\epsilon_n\phi_n(\mathbf r)$. Define $\xi_n=\epsilon_n-\mu$, with $\mu = E_{\mathrm{F}} + \min_n\epsilon_n$ so that $E_{\mathrm{F}}$ is measured from the bottom of the band. For a uniform local induced pairing $\Delta^{\mathrm{ind}}$, orthogonality of the normal-state orbitals makes the pairing diagonal in this basis. Each orbital therefore produces an independent BdG block,
\begin{eqnarray}
H_n
=
\xi_n\tau^z+\Delta^{\mathrm{ind}}\tau^x,
\qquad
E_n
=
\sqrt{\xi_n^2+(\Delta^{\mathrm{ind}})^2}.
\label{eq: modes BdG}
\end{eqnarray}
The corresponding real coherence factors are
\begin{eqnarray}
u_n^2
=
\frac12\left(1+\frac{\xi_n}{E_n}\right),
\qquad
v_n^2
=
\frac12\left(1-\frac{\xi_n}{E_n}\right).
\label{eq: coherence factors results}
\end{eqnarray}

In terms of these normal modes, the normal and anomalous second-order coefficients take the form
\begin{subequations}
\begin{eqnarray}
\frac{K_{ij}^{(pq)}}{|\gamma_{i,p}\gamma_{j,q}|}
&=&
\sum_{n,m}
F_{nm}^{i,p}F_{nm}^{j,q}
\frac{v_n^2v_m^2+u_n^2 u_m^2}{E_n+E_m}, \label{eq: unit envelope K results}
\\
\frac{C_{ij}^{(pq)}}{|\gamma_{i,p}\gamma_{j,q}|}
&=&2
\sum_{n,m}
F_{nm}^{i,p}F_{nm}^{j,q}
\frac{u_nu_mv_nv_m}{E_n+E_m}.
\label{eq: unit envelope C results}
\end{eqnarray}
\end{subequations}
The form factors $F_{nm}^{i,p}$ are determined by the overlap of the contact envelopes $f_{i,p}(\mathbf r)$ with the normal-state wave functions $\phi_n(\mathbf r)$ and $\phi_m(\mathbf r)$ [see Eq.~\eqref{eq: contact form factor methods} for defining equation]. The invariance of the coefficients above under $(u_n,v_n)\mapsto(v_n,-u_n)$ reflects the particle--hole spectral symmetry $\Xi=\tau^y$, with $\Xi \,H_n\,\Xi^\dagger=-H_n$ for all modes $n$.

The sums in Eq.~\eqref{eq: unit envelope K results} and \eqref{eq: unit envelope C results} are effectively restricted to orbitals within the superconducting window,
\begin{eqnarray}
|\xi_n|\lesssim\Delta^{\mathrm{ind}},
\end{eqnarray}
since modes far outside this window are suppressed by their coherence factors and quasiparticle-energy denominators. For a short superconductor $L_x\lesssim\hbar\pi/\sqrt{2m^\star\Delta^{\mathrm{ind}}}$, the normal-state level spacing is comparable to or larger than the induced superconducting gap. Consequently, near the band edge, the mode sums are dominated by the lowest-energy states.

For the I-gadget, this manifold is a nearly degenerate mirror-symmetry doublet (Fig.~\ref{fig: H lowest parity modes}a,b). In the limit of decoupled legs (removing the coupling to the bridge), the lowest states localized on the top and bottom legs are degenerate. The bridge weakly hybridizes them into symmetric and antisymmetric combinations, producing a splitting that is exponentially small in the I-gadget height and much smaller than $\Delta^{\mathrm{ind}}$. Neglecting this splitting, both modes satisfy $\xi_\pm\simeq-E_{\mathrm{F}}$, and at the very edge of the band $E_{\mathrm{F}}\simeq 0$,
\begin{eqnarray}
u_\pm\simeq v_\pm\simeq\frac{1}{\sqrt2}.
\end{eqnarray}
Inspection of Eqs.~\eqref{eq: unit envelope K results} and \eqref{eq: unit envelope C results} then shows that equal electron and hole coherence factors make the pair-addition and pair-removal matrix elements approximately equal in magnitude, yielding
\begin{eqnarray}
K_{ij}^{(pq)}\simeq C_{ij}^{(pq)}
\end{eqnarray}
within the lowest-doublet approximation. Finite splitting and higher modes entering the superconducting window generate small corrections to this equality.

\begin{figure}[t]
    \centering
    \includegraphics[width=0.8\linewidth]{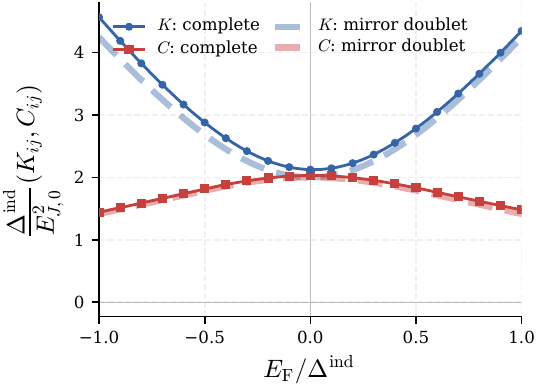}
\caption{Normalized exchange and anomalous pair-transfer amplitudes near the band edge. The dimensionless coefficients $\Delta^{\mathrm{ind}}K_{ij}/E_{J,0}^2$ (blue circles) and $\Delta^{\mathrm{ind}}C_{ij}/E_{J,0}^2$ (red squares) are shown as functions of $E_{\mathrm F}/\Delta^{\mathrm{ind}}$. Solid curves include all normal-state modes up to a cutoff, while the broad dashed curves show the analytical mirror-doublet approximation.}
\label{fig:h_EJ0_normalized}
\end{figure}

This result also follows from a simple Nambu-space exchange symmetry, distinct from the previous particle-hole $\Xi$. The unitary transformation $\mathcal S=\tau^x$ exchanges the electron and hole components and maps
\begin{eqnarray}
\mathcal S H_n(\xi_n)\mathcal S^{\dagger}
=
H_n(-\xi_n).
\label{eq: Nambu exchange symmetry}
\end{eqnarray}
The transformation exchanges $u_n\leftrightarrow v_n$ and interchanges pair addition and removal. In the degenerate band-edge limit, $\xi_\pm=0$, the two lowest BdG blocks are invariant under this exchange, and the normal and anomalous responses coincide.

To remove the arbitrary scalar normalization of each contact profile, we use the local Josephson energies $E_{J,i,p}$ defined in Eq.~\eqref{eq: local Josephson energy} within the lowest-doublet projection. For equal contact energies, $E_{J,i,p}=E_{J,0}$, the lowest-doublet modes yield $K^{(tb)}= K^{(bt)}\simeq0$ and $C^{(tb)}= C^{(bt)}\simeq0$, while the total responses are
\begin{subequations}
\begin{align}
\frac{\Delta^{\mathrm{ind}}K_{ij}}{E_{J,0}^{2}}
&\simeq
\frac{
2\left[
1+2\left(E_{\mathrm F}/\Delta^{\mathrm{ind}}\right)^2
\right]
}{
\sqrt{
1+\left(E_{\mathrm F}/\Delta^{\mathrm{ind}}\right)^2
}
},\label{eq: lowest doublet responses main K}
\\[4pt]
\frac{\Delta^{\mathrm{ind}}C_{ij}}{E_{J,0}^{2}}
&\simeq
\frac{2}{
\sqrt{
1+\left(E_{\mathrm F}/\Delta^{\mathrm{ind}}\right)^2
}
}.\label{eq: lowest doublet responses main C}
\end{align}
\end{subequations}
 At the band edge $E_{\mathrm F}=0$, we have the equality $K_{ij} \simeq C_{ij}$.

Figure~\ref{fig:h_EJ0_normalized} compares the analytical lowest-doublet prediction in Eqs.~\eqref{eq: lowest doublet responses main K} and \eqref{eq: lowest doublet responses main C} with the numerical calculation of 
 $K_{ij}$ and $C_{ij}$ in Eqs.~\eqref{eq: unit envelope K results} and \eqref{eq: unit envelope C results} that takes into account all the modes.
Their close agreement on the interval $-1\leq E_{\mathrm F}/\Delta^{\mathrm{ind}}\leq 1$ demonstrates that the
response near the band edge is heavily dominated by the lowest mirror-symmetry doublet,
while the small residual deviations come from contributions of higher
normal-state modes.
To provide analytical intuition for this result, in Appendix~\ref{supp: local pairing slab} we derive the corresponding coefficients for a quasi-one-dimensional wire and show how the mode expansion converges when the longitudinal level spacing exceeds the induced superconducting gap.

\begin{figure}[t]
    \centering
    \includegraphics[width=0.8\linewidth]{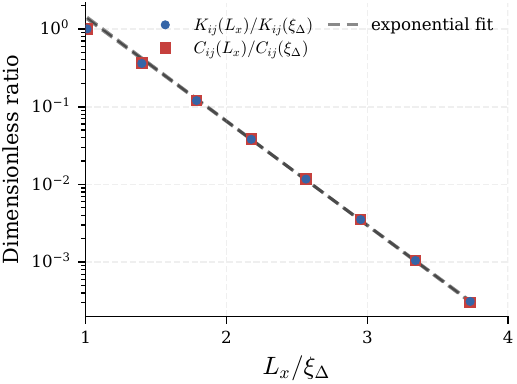}
    \caption{Band-edge length dependence of the I-gadget. The dimensionless, contact-normalized responses $K_{ij}(L_x)/K_{ij}(\xi_\Delta)$ and $C_{ij}(L_x)/C_{ij}(\xi_\Delta)$ are shown as functions of $L_x/\xi_{\Delta}$. Every normal mode is included at each length, as lowest-mode approximation no longer holds.}
    \label{fig: H length scaling}
\end{figure}

At the normal-state band edge, the intrinsic single-quasiparticle decay length is
\begin{eqnarray}
\xi_{\Delta}
=
\frac{\hbar}{\sqrt{m^\star\Delta^{\mathrm{ind}}}},
\label{eq: edge coherence length}
\end{eqnarray}
 fixed by the induced gap and the band-edge effective mass. The scale $\xi_{\Delta}$ follows from the complex-momentum poles of the zero-frequency BdG propagator at $E_{\mathrm F}=0$. For sufficiently long wires, the lowest-mode approximation breaks down and many longitudinal modes contribute to the response. In this regime, each virtual quasiparticle acquires an envelope $e^{-L_x/\xi_{\Delta}}$, so the pair amplitudes decay asymptotically as $K_{ij},C_{ij}\propto e^{-2L_x/\xi_{\Delta}}$,
up to algebraic and contact-dependent prefactors. This scaling is confirmed numerically in Fig.~\ref{fig: H length scaling}, where $K_{ij}$ and $C_{ij}$ exhibit nearly identical effective decay exponents at $E_{\mathrm F}=0$ even after higher-energy modes are included. A complementary analytical discussion for a quasi-one-dimensional wire is given in Appendix~\ref{supp: local pairing slab}.

To quantify deviations from the bond-directional limit, we define the ratio of the unwanted to desired bond components,
\begin{eqnarray}
\frac{\alpha_{ij}}{J_{ij}}
\equiv
\frac{K_{ij}-C_{ij}}{K_{ij}+C_{ij}}\simeq\frac{\left(E_{\mathrm{F}} / \Delta^{\mathrm{ind }}\right)^2}{1+\left(E_{\mathrm{F}} / \Delta^{\mathrm{ind }}\right)^2} .
\end{eqnarray}
The point $\alpha_{ij}/J_{ij}=0$ corresponds to the Kitaev limit, while $\alpha_{ij}/J_{ij}\neq0$ quantifies the residual $\hat\sigma_i^{\mathrm y}\hat\sigma_j^{\mathrm y}$ ($\hat\sigma_i^{\mathrm x}\hat\sigma_j^{\mathrm x}$) interaction on $\mathrm x$ ($\mathrm y$) edges. As one deviates from $E_{\mathrm F}=0$, controllable perturbations $\alpha_{ij}/J_{ij}\simeq (E_{\mathrm F}/\Delta^{\mathrm{ind}})^2$ are taken into account.

To leading order, the low-energy Hamiltonian of the full network is a sum of the local link interactions in Eqs.~\eqref{eq: insulating strip}, \eqref{eq: effective_SC x}, and~\eqref{eq: effective_SC y}, in which processes involving multiple links enter only at higher order. We therefore organize the effective Hamiltonian as
\begin{eqnarray}
    \widehat H_{\mathrm{eff}}
    =
    \widehat H_{\mathrm{Kitaev}}
    +\delta\widehat H_{K-C}
    +\delta\widehat H_{\gamma,g}
    +\delta\widehat H_{\mathrm{lr}}
    +\delta\widehat H_{\mathrm{ho}},
\end{eqnarray}
where the correction terms quantify departures from the ideal Kitaev limit.

The first two corrections arise already at the level of individual links. The term $\delta\widehat H_{K-C}$ contains the unwanted transverse bond components generated by an imbalance between $\mathrm{ECT}^2$ and $\mathrm{CAR}^2$, quantified by $\alpha_{ij}/J_{ij}$. The term $\delta\widehat H_{\gamma,g}$ contains residual single-spin fields arising from asymmetries between the top and bottom contacts and from charge offsets. Contact asymmetries generate transverse fields $g_i$, while charge offsets produce longitudinal fields $g_i^z$ that can be compensated by gate voltages. Provided these fields remain perturbative relative to the Kitaev couplings, they do not invalidate the effective description. Moreover, controlled single-spin fields can be exploited to gap the Majorana spectrum and access the non-Abelian phase. Quantitative bounds on these perturbations are discussed in the Experimental Considerations.

The remaining corrections arise beyond the nearest-neighbor link description. The term $\delta\widehat H_{\mathrm{lr}}$ describes couplings between non-neighboring boxes mediated by the connected superconducting network. These interactions are exponentially suppressed with propagation distance, with a characteristic scale $J_{\mathrm{lr}}\sim J_{\mathrm{x},\mathrm{y}}\, e^{-2L_y/\xi_{\Delta}}$. Finally, $\delta\widehat H_{\mathrm{ho}}$ collects higher-order processes involving multiple links, which are suppressed both by powers of $E_J/\Delta^{\mathrm{ind}}$ and by exponentially small propagation factors set by $e^{-2L_x/\xi_{\Delta}}$. 
These parametric hierarchies establish the Kitaev Hamiltonian as the leading low-energy description without requiring the correction terms to vanish exactly.

\subsection{Experimental considerations}
\label{sec: exp considerations}

The proposed brick-wall-like layout is naturally compatible with planar fabrication and can be implemented in a hybrid Al/InAs heterostructure, where a two-dimensional electron gas (2DEG) forms at the InAs interface and acquires superconducting pairing through proximity to an epitaxial Al film. In such heterostructures, the induced gap typically satisfies $0.33<\Delta^{\mathrm{ind}}/\Delta_{\mathrm{Al}}<0.67$~\cite{Microsoft2023InAs}, with $\Delta_{\mathrm{Al}}\simeq160$--$370\,\mu\mathrm{eV}$. We therefore take $\Delta^{\mathrm{ind}}=200\,\mu\mathrm{eV}$ as a representative value. Josephson junction energies range $10-1000\, \mu \mathrm{eV}$. For the estimates below, we use $E_J=100\,\mu\mathrm{eV}$. Electrostatic gates tune the InAs 2DEG close to the bottom of its band, while the extended Al film provides a superconducting condensate with negligible charging energy. We assume operation in a dilution refrigerator at temperatures $T\sim10$--$20\,\mathrm{mK}$, corresponding to a thermal energy $k_{\mathrm B}T\sim1$--$2\,\mu\mathrm{eV}$.

\begin{figure}[t]
    \centering
    \includegraphics[width=1.0\linewidth]{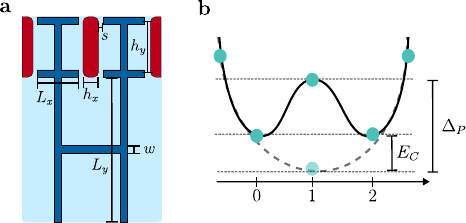}
    \caption{\textbf{a} Geometric dimensions of the superconducting network. The I-gadgets are vertically separated by $L_y$ from each other and have horizontal length $L_x$, and width $w$, while the Cooper-pair boxes have characteristic size $h_x\times h_y$. The gap between the CPB and I-gadgets is captured by $s$. \textbf{b} Effective energy $E_{\mathrm{box}}(n)$ spectrum of a CPB with $n$ electrons in the presence of a parity gap $\Delta_P$.}
    \label{fig: charging energy of CPB}
\end{figure}

For the CPBs, parity effects have been observed across a broad range of superconducting-island sizes, from nanoscale grains~\cite{Black1996scgap,Ralph1995electronic,Ralph1996metal} up to micrometer-scale islands~\cite{Tuominen1997parity,Lafarge1993parity}. The associated parity gap is of the order of the superconducting gap, with experimentally reported values in the range $0.02$--$0.3\,\mathrm{meV}$. Here we take it to be $\Delta_P\simeq200\,\mu\mathrm{eV}$.  For our estimates (see Sec.~\ref{sec: 
further 
details}), we model each proximitized island as a rectangle of sides $ h_x \simeq50\,\mathrm{nm}$ and $h_y\simeq 1\,\mu\mathrm{m}$ (Fig.~\ref{fig: charging energy of CPB}a,b). This geometry yields a charging energy $E_C\simeq45\,\mu\mathrm{eV}$, allowing the required regime $\Delta_P>E_C$ to be reached (see Sec.~\ref{sec: 
further 
details} for details).

With $m^\star=0.023m_e$ for InAs, Eq.~\eqref{eq: edge coherence length} gives $\xi_{\Delta}\simeq129\,\mathrm{nm}$. We therefore consider I-gadgets with intermediate lengths $L_x=100-300\,\mathrm{nm}$, comparable to $\xi_\Delta$. These dimensions are compatible with current epitaxial Al/InAs fabrication capabilities~\cite{Microsoft2025interferometric}. At the band edge, our numerical results give $\Delta^{\mathrm{ind}}K_{ij}/E_J^2\simeq0.8-2.1$, corresponding to $J_{\mathrm{x},\mathrm{y}}\simeq40-105\,\mu\mathrm{eV}$. Reaching the isotropic point further requires the capacitive $\mathrm z$-bond coupling to be engineered to the same scale, which corresponds to $L_y\simeq3.6-9.5\,\mu\mathrm{m}$.

The required magnetic fluxes further constrain the device geometry. For a uniform perpendicular field $H$, the conditions $\Phi_A=\Phi_0/2\ \mathrm{mod}\ \Phi_0$ and $\Phi_B=\Phi_0/4\ \mathrm{mod}\ \Phi_0$ require
\begin{align}
\Phi_A
=
\left(\frac{L_x-w}{2}+s\right)\left(h_y-w\right)\, H
=
\left(n_A+\frac{1}{2}\right)\Phi_0,
\nonumber\\
\Phi_B
=
\left(L_x-w+h_x+2s\right)\left(\frac{L_y-w}{2}\right) H
=
\left(n_B+\frac{1}{4}\right)\Phi_0,
\end{align}
where $n_A,n_B\in\mathbb{Z}$ specify the integer number of flux quanta threading the corresponding loops.
For the geometry considered here, $n_A=0$ and $n_B=2$ allow the two flux
conditions to be made commensurate through small adjustments of the loop
dimensions, yielding experimentally accessible fields of order
$H\simeq 10\,\mathrm{mT}$, below the characteristic critical-field
scale of Al $H\simeq 0.1\, \mathrm T.$

At approximately isotropic couplings, the device realizes the gapless $B$ phase of the Kitaev model~\cite{Kitaev2006exactly}, which is the primary target of our construction. For the estimated range, the vison gap $\Delta_v=0.2633J_{\mathrm{x},\mathrm{y}}$~\cite{Panigrahi2023gap} is $\Delta_v\simeq10-28\,\mu\mathrm{eV}$, above the proposed $1-2\,\mu\mathrm{eV}$ thermal scale.

As a prospective extension, controlled single-spin fields $g_i$ and $g^z_i$ could be used to gap the Majorana spectrum and access the non-Abelian phase. Such fields may be engineered through charge biases, contact asymmetries, or controlled flux deviations, with a scale set by the local Josephson energies; for example, $|g_i|\propto E_{J}|\delta\Phi|/\Phi_0$ for small flux detuning. When all three spin components are nonzero, third-order perturbation theory generates a Majorana mass $\Delta_M\simeq 5\sqrt 3 |g_xg_yg_z|/J^2$~\cite{Kitaev2006exactly}, which can be tuned to a scale comparable to the vison gap $\Delta_v$. For $J\simeq100\,\mu\mathrm{eV}$, this estimate requires fields of magnitude $|g_a|\simeq30\,\mu\mathrm{eV}$.  The presence of such couplings breaks time-reversal symmetry together with the relevant combined antiunitary symmetry involving time reversal and the corresponding lattice reflection.

\section{Further Details}
\label{sec: 
further 
details}

\subsection{Modeling of Cooper-pair boxes}

We model each superconducting vertex as a Cooper-pair box (CPB) with electron-number $n$. The energy of an isolated box is approximated by~\cite{vanHeck2016conductance} 
\begin{eqnarray}
E_{\mathrm{box}}(n)
=
E_C(n-n_g)^2
+
\Delta_P\,\frac{1-(-1)^{n}}{2},
\label{eq: charging energy}
\end{eqnarray}
where $E_C=e^2/2C$ is the charging energy, $n_g$ is the gate-induced offset charge, and $\Delta_P$ is the parity gap (Fig.~\ref{fig: charging energy of CPB}b). We operate at $n_g=1$, for which the charging contribution alone favors the odd state $n=1$, while the parity gap raises it relative to the even sector. The states $n=0$ and $n=2$ are then degenerate with energy $E_C$, while the $n=1$ state is separated from this two-dimensional subspace by $\Delta_{\mathrm{odd}}=\Delta_P-E_C$.

To estimate the charging energy, we take the dominant capacitance of each CPB to arise from a nearby metallic layer separated from the superconducting island by a thin HfO$_2$ dielectric of thickness $d_0\simeq5\,\mathrm{nm}$. Approximating the island and metallic layer as parallel plates with overlap area $h_xh_y$, the capacitance is
\begin{equation}
C
\simeq
\epsilon_0\epsilon_{\mathrm{HfO_2}}
\frac{h_xh_y}{d_0},
\end{equation}
where $\epsilon_{\mathrm{HfO_2}}\simeq16$--$25$ is the relative permittivity of HfO$_2$~\cite{Gritsenko2016Hafnium}. Taking $\epsilon_{\mathrm{HfO_2}}\simeq20$, $h_x\simeq50\,\mathrm{nm}$, and $h_y\simeq1\,\mu\mathrm{m}$ gives $E_C\simeq45\,\mu\mathrm{eV}$. The CPB dimensions are chosen to be compatible with the flux conditions imposed on the $\Phi_A$ and $\Phi_B$ loops. Additional capacitances associated with the contacts between the CPB and neighboring I-gadgets provide subleading corrections and are neglected here.

The parity gap raises the odd-parity sector relative to the even sector, leaving the $n=0$ and $n=2$ states degenerate with energy $E_C$. For $\Delta_P\simeq200\,\mu\mathrm{eV}$ and the charging energy estimated above, the $n=1$ state is separated from this two-dimensional low-energy subspace by
\begin{equation}
\Delta_{\mathrm{odd}}
=
\Delta_P-E_C
\simeq155\,\mu\mathrm{eV}.
\end{equation}
 The states $n=0$ and $n=2$ therefore form a well-isolated low-energy doublet defining the effective spin-$1/2$ degree of freedom. We further note that the CPBs are chosen to be sufficiently small that the magnetic flux threading each island remains well below a single flux quantum.

We now make the effective spin representation explicit. Let $\hat d_{i\sigma}^\dagger$ and $\hat d_{i\sigma}$ create and annihilate an electron with spin $\sigma=\uparrow,\downarrow$ on CPB $i$, and define $\hat n_{i}=\hat d_{i\uparrow}^\dagger\hat d_{i\uparrow}+\hat d_{i\downarrow}^\dagger\hat d_{i\downarrow}$. Within the even-parity subspace, the spin operators are represented as
\begin{align}
\hat\sigma_i^+
&=
e^{-\mathrm{i}\varphi}\,
\hat d_{i\uparrow}^\dagger\hat d_{i\downarrow}^\dagger,
&
\hat\sigma_i^-
&=
e^{\mathrm{i}\varphi}\,
\hat d_{i\downarrow}\hat d_{i\uparrow},
&
\hat\sigma_i^z
&=
\hat n_{i}-1,
\label{map_explicit}
\end{align}
with $\hat\sigma_i^\pm=(\hat\sigma_i^{\mathrm{x}}\pm\mathrm{i}\hat\sigma_i^{\mathrm{y}})/2$. Acting on the basis $\{|n=0\rangle,|n=2\rangle\}$, these operators obey the $\mathrm{SU}(2)$ algebra, with the empty and doubly occupied states identified with the two spin states. The phase $\varphi\in[0,2\pi)$ reflects the $U(1)$ freedom in defining the pair operators and can be chosen to render the superconducting order parameter real~\footnote{Equivalently, $\varphi$ fixes the phase convention for the effective spin raising and lowering operators. We henceforth choose a gauge in which the superconducting order parameter is real\label{alpha_parameter}}.

\subsection{Numerical modeling of I-gadgets}

We model the proximitized semiconductor with a spatially uniform onsite spin-singlet pairing $\Delta^{\mathrm{ind}}$ over the I-shaped geometry $\mathrm H$ between CPBs $i$ and  $j$ via
\begin{eqnarray}
    \widehat H_{ij}^{\mathrm{SC}}
=
\int_{\mathrm H} d^2\mathbf r\,
\widehat\Psi^\dagger(\mathbf r)
\,\mathcal H_{\mathrm{BdG}}(\mathbf r)\,
\widehat\Psi(\mathbf r),
\end{eqnarray}
with
\begin{eqnarray}
\mathcal H_{\mathrm{BdG}}(\mathbf r)
&=&
\begin{pmatrix}
h_0-\mu & \Delta^{\mathrm{ind}}\\
\Delta^{\mathrm{ind}} & -(h_0-\mu)
\end{pmatrix}
,
\qquad
h_0=-\frac{\hbar^2\nabla^2}{2m^\star},
\label{eq: bdg Hamiltonian}
\end{eqnarray}
in the Nambu basis $\widehat \Psi(\mathbf r) = \left[\hat\psi_\uparrow(\mathbf r),\hat\psi_\downarrow^\dagger(\mathbf r)\right]^{\scriptscriptstyle T}$~\footnote{We omit the explicit minimal coupling to the vector potential and instead choose a gauge in which the flux dependence is encoded in the Peierls phases of the couplings \(\gamma_{i,p}\), as illustrated in Fig.~\ref{array}d,e}.
    Let $h_0\phi_n=\epsilon_n\phi_n$ with real orthonormal normal-state modes. Let $\epsilon_0$ be the energy of the lowest mode.  We set $\mu=\epsilon_0+E_{\mathrm F}$, so that $\xi_n=\epsilon_n-\epsilon_0-E_{\mathrm F}$. The electron annihilation operator of spin $\sigma$ in orbital $n$ is
\begin{eqnarray}
    \hat\psi_{n\sigma}
=
\int d\mathbf r\,
\phi_n^\ast(\mathbf r)\hat\psi_\sigma(\mathbf r),
\end{eqnarray}
which induces the BdG Hamiltonian in the mode basis
\begin{eqnarray}
H_n = \begin{pmatrix}
\xi_n & \Delta^{\mathrm{ind}}\\
\Delta^{\mathrm{ind}} & -\xi_n
\end{pmatrix}    ,
\end{eqnarray}
 corresponding to the independent blocks in Eq.~\eqref{eq: modes BdG}.

In order to diagonalize $H_n$, we introduce  the canonical Bogoliubov transformation
\begin{subequations}
\begin{eqnarray}
\hat \psi_{n\uparrow}
&=&u_n\hat\gamma_{n\uparrow}
-v_n\hat\gamma_{n\downarrow}^\dagger,\\
\hat \psi_{n\downarrow}
&=&u_n\hat\gamma_{n\downarrow}
+v_n\hat\gamma_{n\uparrow}^\dagger.
\end{eqnarray}
\label{eq: canonical Bogoliubov transformation}
\end{subequations}
in terms of the nonnegative coherence factors in Eq.~\eqref{eq: coherence factors results}
\begin{align}
u_n^2=\frac12\left(1+\frac{\xi_n}{E_n}\right),
\quad
v_n^2=\frac12\left(1-\frac{\xi_n}{E_n}\right),
\quad
u_nv_n=\frac{\Delta^{\mathrm{ind}}}{2E_n}.
\label{eq: coherence factors methods}
\end{align}

For the $i-th$ contact on leg $p$ of I-gadget, we define the envelope form factors present in Eq.~\eqref{eq: unit envelope K results} and \eqref{eq: unit envelope C results}
\begin{eqnarray}
F_{nm}^{i,p}
=
\int d^2\mathbf r\,
 f_{i,p}(\mathbf r)\phi_n(\mathbf r)\phi_m(\mathbf r).
\label{eq: contact form factor methods}
\end{eqnarray}
For $|n\uparrow,m\downarrow\rangle=\hat\gamma_{n\uparrow}^\dagger\hat\gamma_{m\downarrow}^\dagger|\Omega\rangle$, the distinct pair-addition and pair-removal amplitudes at unit envelope normalization are
\begin{subequations}
\begin{eqnarray}
\langle n\uparrow,m\downarrow|\hat b_{i,p}^\dagger|\Omega\rangle
&=&F_{nm}^{i,p}\, u_nu_m,\\
\langle n\uparrow,m\downarrow|\hat b_{i,p}|\Omega\rangle
&=&F_{mn}^{i,p}\,  v_nv_m.
\end{eqnarray}
\label{eq: addition removal methods}
\end{subequations}
 These allow us to write both $K_{ij}$ and $C_{ij}$ in terms of the coherence factors and lead to Eq.~\eqref{eq: unit envelope K results} and \eqref{eq: unit envelope C results}.
Similarly, the local condensate factor in Eq.~\eqref{eq: local Josephson energy} can be written in terms of the coherence factors
\begin{eqnarray}
E_{J,i,p} 
=|\gamma_{i,p}|
\sum_nF_{nn}^{i,p}\frac{\Delta^{\mathrm{ind}}}{2E_n},
\label{eq: contact condensate methods}
\end{eqnarray}
which is used to eliminate the bare couplings $|\gamma_{i,p}|$.

\begin{figure}
    \centering
    \includegraphics[width=0.8\linewidth]{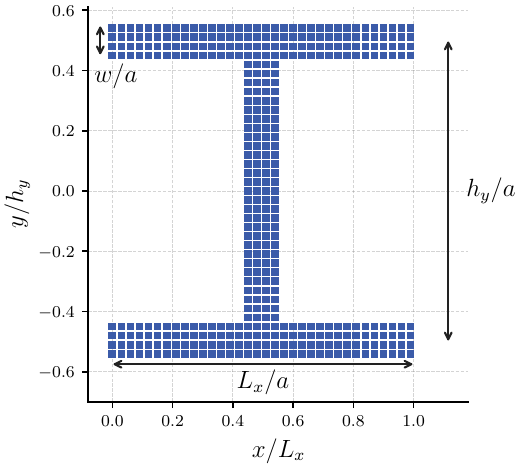}
    \caption{Numerical  discretization of I-gadget. The horizontal legs have length $L_x$ and center-to-center separation $h_y$, all segments have width $w$, and the square-lattice spacing is $a$.}
    \label{fig: discretization H gadget}
\end{figure}

For the numerical calculation, we discretize the I-shaped semiconductor region on a square lattice with spacing $a$, as illustrated in Fig.~\ref{fig: discretization H gadget}. The finite-difference kinetic energy has onsite elements $4t$ and nearest-neighbor hopping $-t$, with $t=\hbar^2/(2m^\star a^2)$ and hard-wall boundary conditions. 

The coupling between a CPB and an I-gadget is modeled by a spatial envelope localized near each endpoint. For $p=t,b$, labeling the top and bottom legs, we take exponential profiles localized near the left ($L$) and right ($R$) ends of the gadget,
\begin{eqnarray}
 f_{L,p}(x,y)
&=& A\, 
\Theta_p(y)e^{-x/\ell_c},
\nonumber\\
 f_{R,p}(x,y)
&=& A\, 
\Theta_p(y)e^{-(L_x-x)/\ell_c},
\end{eqnarray}
where
\begin{eqnarray}
\Theta_p(y)
=
\begin{cases}
1, & |y-y_p|\leq w/2,\\
0, & \mathrm{otherwise},
\end{cases}
\end{eqnarray}
with $y_t=+h_y/2$ and $y_b=-h_y/2$ (see Fig.~\ref{fig: discretization H gadget}), and normalization factor
\begin{eqnarray}
   A= \frac{\sqrt{2}}{\sqrt{w\, \ell_c \left(1-e^{-\frac{2 L_x}{\ell_c}}\right)}}.
\end{eqnarray}

The calculations in Fig.~\ref{fig:h_EJ0_normalized} use
\begin{eqnarray}
&&a=5\,\mathrm{nm},\quad L_x=130\,\mathrm{nm},\quad
h_y=1000\,\mathrm{nm},\quad w=50\,\mathrm{nm},\nonumber\\
&&\ell_c=40\,\mathrm{nm},\quad m^\star=0.023m_e,
\quad\Delta^{\mathrm{ind}}=0.2\,\mathrm{meV}.
\label{eq: local pairing numerical parameters}
\end{eqnarray}

 The solid H curves in Fig.~\ref{fig:h_EJ0_normalized} take into account all modes below a Debye energy cutoff $\Lambda\simeq 10\, \mathrm{meV}$. For the large bridge length $h_y$ considered here, the splitting of the lowest-energy doublet is negligible $ 4.58\times10^{-9}\Delta^{\mathrm{ind}}$, justifying its approximation to zero. Similarly, the results in Fig.~\ref{fig: H length scaling} also take into account all modes and use parameters comparable to those above, with $E_{\mathrm F}=0$ fixed while varying $1\leq L_x/\xi_{\Delta}\leq 4$. Evaluating $E_J$ from Eq.~\eqref{eq: contact condensate methods}, we find $E_J\simeq\eta E_{J,0}$, where $\eta$ is an overall $\mathcal O(1)$ renormalization factor incorporated into the estimates in the experimental considerations (Sec.~\ref{sec: exp considerations}).

\section{Conclusion}
\label{sec: discussion}

In this work, we have proposed a route to realizing the Kitaev honeycomb model using a planar network of superconducting Cooper-pair boxes coupled through semiconductor--superconductor I-gadgets. The central ingredient is the conversion of conventional superconducting processes into bond-directional spin interactions. Capacitive coupling produces $\hat\sigma^{\mathrm z}\hat\sigma^{\mathrm z}$ interactions along the $\mathrm z$ bonds, while the interplay of elastic pair cotunneling
and double crossed Andreev reflection processes combine to produce $\hat\sigma^{\mathrm x}\hat\sigma^{\mathrm x}$ and $\hat\sigma^{\mathrm y}\hat\sigma^{\mathrm y}$ interactions along the $\mathrm x$ and $\mathrm y$ bonds, respectively. We showed that the required balance between these processes, $K_{ij}\simeq C_{ij}$, emerges naturally near the bottom of the semiconductor band from an approximate Nambu-exchange symmetry. The numerical results for the I-gadget consistently identify a finite operating window around this point. These results establish a controlled regime in which the Kitaev Hamiltonian provides the leading low-energy description of the network.

Our estimates for hybrid Al/InAs devices focus primarily on the gapless and non-Abelian regimes, placing the required dimensions, magnetic fields, and interaction strengths within experimentally relevant ranges while maintaining the characteristic Kitaev energy scales above the proposed operating temperature. In particular, the approximately isotropic couplings realize the gapless Kitaev spin liquid, while controlled local fields provide a route to gap the Majorana spectrum and access the non-Abelian phase. More broadly, our approach illustrates how geometry, magnetic flux, and superconducting interference can be combined to engineer highly anisotropic spin interactions from standard ingredients of hybrid superconducting circuits. The same design principles may be extended to other lattices and interaction patterns, providing a framework for engineering frustrated and topological quantum spin models in planar semiconductor--superconductor architectures.





\section*{Acknowledgments}


{The work of G.D., M.D., M.J.M., C.M.M. and C.C. is supported by the U.S. Department of Energy, Office of Basic Energy Sciences, under Award DE-SC0026189.} 




\appendix

\begin{widetext}

\section{Analytical derivation for a thin superconducting wire}
\label{supp: local pairing slab}

The full I-shaped geometry requires a numerical diagonalization of the normal-state Hamiltonian. Here we develop an analytical benchmark for a thin superconducting wire (so that transverse excitations may be neglected), modeled directly as a one-dimensional hard-wall segment $0<x<L_x$. Its normalized normal-state eigenfunctions are
\begin{equation}
\phi_n(x)=\sqrt{\frac{2}{L_x}}
\sin\left(\frac{n\pi x}{L_x}\right),
\qquad n=1,2,\ldots,
\end{equation}
with detunings
\begin{equation}
\xi_n=\delta_x(n^2-1)-E_{\mathrm F},
\qquad
\delta_x=\frac{\hbar^2\pi^2}{2m^\star L_x^2},
\label{eq: supplement slab detunings}
\end{equation}
where $E_{\mathrm F}=0$ places the chemical potential at the bottom of the confined band. The spacing between the two lowest longitudinal levels is $3\delta_x$.

It is convenient to introduce the dimensionless quantities
\begin{equation}
e\equiv\frac{E_{\mathrm F}}{\Delta^{\mathrm{ind}}},
\qquad
r\equiv\frac{3\delta_x}{\Delta^{\mathrm{ind}}},
\qquad
x_n\equiv\frac{\xi_n}{\Delta^{\mathrm{ind}}}
=\frac{r}{3}(n^2-1)-e,
\qquad
q_n\equiv\frac{E_n}{\Delta^{\mathrm{ind}}}
=\sqrt{1+x_n^2}.
\label{eq: slab dimensionless variables}
\end{equation}
For uniform local pairing, each normal orbital produces an independent BdG block, with coherence factors
\begin{equation}
u_n^2=\frac{q_n+x_n}{2q_n},
\qquad
v_n^2=\frac{q_n-x_n}{2q_n},
\qquad
u_nv_n=\frac{1}{2q_n}.
\label{eq: slab dimensionless coherence}
\end{equation}

\subsection{Contact form factors and exact mode sums}

We place contacts at the left and right ends of the wire, denoted by $s=L,R$, and take equal scalar amplitudes,
\begin{equation}
f_L(x)=A e^{-x/\ell_c},
\qquad
f_R(x)=A e^{-(L_x-x)/\ell_c}, \quad  A=\frac{\sqrt{2}}{\sqrt{\ell_c \left(1-e^{-\frac{2 L_x}{\ell_c}}\right)}}.
\end{equation}
Their form factors are
\begin{equation}
F_{nm}^{s}
=
\int_0^{L_x}dx\,f_s(x)\phi_n(x)\phi_m(x),
\end{equation}
and mirror symmetry gives $F_{nm}^{R}=(-1)^{n+m}F_{nm}^{L}$. Writing $a_c=L_x/\ell_c$, the left-contact form factor is
\begin{equation}
F_{nm}^{L}
=
A\,a_c
\left[
1-(-1)^{n+m}e^{-a_c}
\right]
\left[
\frac{1}{a_c^2+\pi^2(n-m)^2}
-
\frac{1}{a_c^2+\pi^2(n+m)^2}
\right].
\label{eq: supplement exponential form factor}
\end{equation}

To display the contribution of each pair of normal modes, we define the complete-ordering kernels
\begin{equation}
\mathcal T_{nm}^{K}
=
\frac{
(q_n+x_n)(q_m+x_m)
+
(q_n-x_n)(q_m-x_m)
}{
4q_nq_m(q_n+q_m)
}
=
\frac{q_nq_m+x_nx_m}
{2q_nq_m(q_n+q_m)},
\qquad
\mathcal T_{nm}^{C}
=
\frac{1}
{2q_nq_m(q_n+q_m)}.
\label{eq: slab kernels}
\end{equation}
The two terms in $\mathcal T_{nm}^{K}$ arise from the pair-addition and pair-removal intermediate states, respectively, while the factor of two in $\mathcal T_{nm}^{C}$ accounts for the two anomalous orderings. Both kernels are invariant under the particle--hole transformation $(u_n,v_n)\mapsto(v_n,-u_n)$.

The propagation amplitudes are then
\begin{equation}
\Delta^{\mathrm{ind}}\mathcal K_{LR}
=\gamma_L \gamma_R
\sum_{n,m}
F_{nm}^{L}F_{nm}^{R}\mathcal T_{nm}^{K},
\qquad
\Delta^{\mathrm{ind}}\mathcal C_{LR}
= \gamma_L \gamma_R
\sum_{n,m}
F_{nm}^{L}F_{nm}^{R}\mathcal T_{nm}^{C}.
\label{eq: slab explicit series}
\end{equation}
The local condensate factor at either contact is
\begin{equation}
P_s
=
\frac{1}{2}
\sum_n\frac{F_{nn}^{s}}{q_n}.
\label{eq: slab explicit P series}
\end{equation}
Using the local Josephson energies $E_{J,L}$ and $E_{J,R}$ at the left and right ends of the wire,
\begin{equation}
K_{LR}
=
\frac{E_{J,L}E_{J,R}}{P_LP_R}\mathcal K_{LR},
\qquad
C_{LR}
=
\frac{E_{J,L}E_{J,R}}{P_LP_R}\mathcal C_{LR},
\label{eq: supplement EJ normalized slab}
\end{equation}
eliminates the bare couplings $\gamma_s$. The normalized spatial profile remains encoded in the form factors.

\subsection{Active modes and lowest-mode regime}

Modes with substantial electron--hole mixing lie within the superconducting window, $|\xi_n|\lesssim\Delta^{\mathrm{ind}}$, or equivalently
\begin{equation}
\left|
\frac{r}{3}(n^2-1)-e
\right|
\lesssim1.
\label{eq: slab active criterion}
\end{equation}
Near the band edge, the number of such modes is approximately
\begin{equation}
N_{\mathrm{act}}
\simeq
\left\lfloor
\sqrt{1+\frac{3}{r}}
\right\rfloor.
\label{eq: slab active number r}
\end{equation}
Since $x_1=-e$ and $x_2=r-e$, the parameter $r$ is exactly the spacing between the two lowest longitudinal levels in units of $\Delta^{\mathrm{ind}}$. Thus, $r\gg1$ defines the spectrally isolated lowest-mode regime, whereas $r\sim1$ marks the crossover at which the second orbital enters the superconducting window.

The same conclusion follows directly from the individual terms of the series. At $e=0$, the lowest kernels are identical,
\begin{equation}
\mathcal T_{11}^{K}
=
\mathcal T_{11}^{C}
=
\frac{1}{4}.
\end{equation}
The leading mixed-mode corrections satisfy
\begin{equation}
\frac{\mathcal T_{12}^{K}}
{\mathcal T_{11}^{K}}
=
\frac{2}{1+q_2}
\sim
\frac{2}{r},
\qquad
\frac{\mathcal T_{12}^{C}}
{\mathcal T_{11}^{C}}
=
\frac{2}{q_2(1+q_2)}
\sim
\frac{2}{r^2},
\qquad
q_2=\sqrt{1+r^2},
\label{eq: slab leading corrections}
\end{equation}
for $r\gg1$. The diagonal $(2,2)$ corrections obey
\begin{equation}
\frac{\mathcal T_{22}^{K}}
{\mathcal T_{11}^{K}}
=
\frac{1+2r^2}{(1+r^2)^{3/2}}
\sim
\frac{2}{r},
\qquad
\frac{\mathcal T_{22}^{C}}
{\mathcal T_{11}^{C}}
=
\frac{1}{(1+r^2)^{3/2}}
\sim
\frac{1}{r^3},
\end{equation}
up to the corresponding contact-form-factor ratios. The complete $K_{LR}$ response is therefore more sensitive than $C_{LR}$ to modes outside the superconducting window. For small wires $L_x\sim \xi_\Delta$, one sees that $r\sim 3\pi^2/2\gg1$, which suppresses higher-modes corrections.

For high-energy modes with $x_n\gg1$,
\begin{equation}
u_n^2
\simeq
1-\frac{1}{4x_n^2},
\qquad
v_n^2
\simeq
\frac{1}{4x_n^2},
\qquad
u_nv_n
\simeq
\frac{1}{2x_n}.
\end{equation}
When both mode indices are large, the kernels consequently behave as
\begin{equation}
\mathcal T_{nm}^{K}
\simeq
\frac{1}{x_n+x_m},
\qquad
\mathcal T_{nm}^{C}
\simeq
\frac{1}
{2x_nx_m(x_n+x_m)}.
\label{eq: slab kernel ultraviolet behavior}
\end{equation}
Thus, the complete $K$ kernel is suppressed only by the energy denominator, whereas the $C$ kernel carries two additional inverse-energy factors.

For the exponential contact profiles considered here, $F_{nm}^{s}$ approaches a finite function of $n-m$ when $n$ and $m$ become large with $n-m$ fixed, while it decays rapidly when the two indices are widely separated. Since $x_n\propto n^2$, the near-diagonal sector determines the asymptotic truncation error for a symmetric cutoff $n,m\leq N$, giving
\begin{equation}
\delta\mathcal K_{LR}(N)
=
\mathcal O(N^{-1}),
\qquad
\delta\mathcal C_{LR}(N)
=
\mathcal O(N^{-5}).
\label{eq: slab ultraviolet tails}
\end{equation}
The local condensate factor converges with the same power as the normal response,
\begin{equation}
\delta P_s(N)=\mathcal O(N^{-1}).
\end{equation}

\subsection{Lowest-mode result}

For $r\gg1$, retaining only $n=m=1$ gives
\begin{equation}
\frac{\Delta^{\mathrm{ind}}K_{LR}^{(1)}}
{E_{J,L}E_{J,R}}
=
\frac{
1+2e^2
}{
\sqrt{1+e^2}
},
\qquad
\frac{\Delta^{\mathrm{ind}}C_{LR}^{(1)}}
{E_{J,L}E_{J,R}}
=
\frac{1}
{\sqrt{1+e^2}}.
\label{eq: slab one mode responses}
\end{equation}
These are the thin-wire analogs of the corresponding mirror-doublet expressions for the I-gadget. Their relative mismatch is
\begin{equation}
\frac{
K_{LR}^{(1)}-C_{LR}^{(1)}
}{
K_{LR}^{(1)}+C_{LR}^{(1)}
}
=
\frac{e^2}{1+e^2}.
\label{eq: supplement one mode mismatch}
\end{equation}
At the band edge, $e=0$, the lowest-mode contributions coincide exactly, $K_{LR}^{(1)}=C_{LR}^{(1)}=1$.

\subsection{Long-wire limit}

The opposite limit $r\ll1$ corresponds to a long wire with a quasi-continuous longitudinal spectrum. In this regime the exponential decay with $L_x$ is not apparent term by term in Eq.~\eqref{eq: slab explicit series}. Rather, it emerges from the coherent sum over the oscillating factors $F_{nm}^{(R)}=(-1)^{n+m}F_{nm}^{(L)}$. Replacing $\pi n/L_x\rightarrow k$ and the mode sums by continuum integrals, the left-to-right propagation is controlled at large $L_x$ by the complex momenta satisfying
\begin{equation}
\frac{\hbar^2k_\star^2}{2m^\star}
=
E_{\mathrm F}\pm i\Delta^{\mathrm{ind}}.
\end{equation}
Writing $k_\star=k_0+i\kappa$, the inverse single-particle decay length is
\begin{equation}
\kappa
=
\sqrt{
\frac{m^\star}{\hbar^2}
\left[
\sqrt{E_{\mathrm F}^2+(\Delta^{\mathrm{ind}})^2}
-E_{\mathrm F}
\right]
}.
\end{equation}
Since both $\mathcal K_{LR}$ and $\mathcal C_{LR}$ involve two quasiparticles propagating between the contacts, their leading asymptotic behavior is
\begin{equation}
\mathcal K_{LR},\,\mathcal C_{LR}
\propto
e^{-2\kappa L_x},
\end{equation}
up to algebraic prefactors. At the band edge, defining
$\xi_\Delta=\hbar/\sqrt{m^\star\Delta^{\mathrm{ind}}}$,
this reduces to
\begin{equation}
\mathcal K_{LR},\,\mathcal C_{LR}
\propto
e^{-2L_x/\xi_\Delta},
\end{equation}
which  justifies the coherence length in Eq.~\eqref{eq: edge coherence length}.

The thin-wire model gives a simple analytical check of the band-edge mechanism discussed in the main text. The accuracy of the lowest-mode approximation is controlled by $r=3\delta_x/\Delta^{\mathrm{ind}}$. For $r\gg1$, contributions from excited longitudinal modes are suppressed, whereas for $r\sim1$, several modes must be retained. At $E_{\mathrm F}=0$, the lowest mode gives $K_{LR}=C_{LR}$ exactly, and the corrections from higher modes follow from the series above.%

\end{widetext}

\section*{References}
\bibliography{ref}

\end{document}